\documentclass{aa}
\usepackage{graphicx}
\begin{document}
   \title{Diffraction-limited CCD imaging with faint reference stars}

   \author{R.N. Tubbs \inst{1} \and J.E. Baldwin \inst{2} \and
C.D. Mackay \inst{1} \and G.C. Cox \inst{3}
          }

   \offprints{R. N. Tubbs}

   \institute{Institute of Astronomy, Madingley Road, Cambridge CB3
              0HA, UK\\
              \email{rnt20@ast.cam.ac.uk}
         \and
             Cavendish Astrophysics Group, Cavendish Laboratory,
              	Madingley Road, Cambridge CB3 0HE, UK\\
             \email{jeb@mrao.cam.ac.uk}
         \and
             Nordic Optical Telescope,
              	Apartado 474,
              	E-38700 Santa Cruz de La Palma,
              	Canarias, Spain\\
             \email{cox@not.iac.es}
             }

   \date{Received March 1, 2002; accepted March 26, 2002}

   \abstract{By selecting short exposure images taken using a CCD with negligible
readout noise we obtained essentially diffraction-limited 810~nm images of faint
objects using nearby reference stars brighter than $I=16$ at a 2.56~m telescope. The FWHM of
the isoplanatic patch for the technique is found to be 50 arcseconds,
providing $\sim$20\% sky coverage around suitable reference
stars.
   }

   \maketitle
%

\section{Introduction}

In recent observations with the Nordic Optical Telescope (NOT) we demonstrated
that selection of the best images from a large dataset of short exposures 
can provide essentially diffraction-limited images at wavelengths shorter
than 1~$\mu$m from well-figured ground-based telescopes as large as 2.5~m
diameter (\cite{baldwin01}). Images of bright stars derived from the best 1\% of 
exposures of 5~ms duration under good seeing had Strehl ratios as high 
as 0.3 at 810~nm. The use of the brightest speckle for image selection, 
shifting and co-adding provides a significant improvement in resolution over
recentring and selection techniques based upon the image centroid
(see \cite{nieto91} and references therein).

The applicability of this technique to faint astronomical 
targets depends on the probability that a sufficiently bright reference star
lies within the isoplanatic patch prevailing at the times of the selected 
exposures. This probability is expected to be very much higher for selected
exposures than for adaptive optics (AO) at the same wavelength for the
following reasons: the 
atmospheric phase variations at the times of the selected exposures are about 
1 rad rms over the whole mirror aperture ($\sim 6-7r_{0}$) rather than
$\sim r_{0}$ for non-conjugate
AO, so the isoplanatic patch is expected to be correspondingly large; 
light 
from the whole aperture ($\sim 6-7r_{0}$) contributes to the short exposure images
rather than $\sim r_{0}$ for AO wavefront sensors; and since selection of exposures is a 
passive activity, exposures can be at least 10 times longer than those 
necessary for servo-correction in AO, further increasing the limiting magnitude.

If the
telescope mirror itself contains phase errors, then the best exposures will
be those where the atmosphere has partially corrected these
errors. The atmospheric phase variations may then
be larger than 1 rad rms over the whole aperture, giving less of an
improvement in isoplanatic patch size over AO. A similar
situation may arise if a good exposure occurs when phase variations
cancel out between two independent atmospheric layers.

A separate consideration is whether the read-out noise of the detector
compromises the quality of the final image obtained by summing the
selected short exposures.
The development of CCDs with negligible readout noise (\cite{mackay01})
has eliminated this noise penalty for fast frame rate imaging. The use
of such CCDs for observations involving image selection should provide
yet further improvement in the limiting magnitude of the reference
star which can be used and also the limiting sensitivity to faint target
objects in the field around the reference star.

The observations described here were designed to test the
performance of the exposure selection technique at low light levels
using a low noise CCD and to investigate the size of the isoplanatic
patch for the technique. We were also
interested in determining the quality of relative
photometry and astrometry within globular cluster images using a range
of reference star magnitudes.
\section{Observations and Data Reduction}
On the nights of 2001 July 5 and 6 we undertook high frame-rate
imaging observations at the Cassegrain focus of the NOT using a 
camera built around a low-noise Marconi CCD. The camera comprised a
front-illuminated $576\times288$
frame-transfer CCD65 with $20\times30$~$\mu$m
pixels cooled in a liquid nitrogen dewar to 120~$\degr$K to
minimise dark current. The CCD was run by an AstroCam 4100 controller
modified to provide a variable voltage clock signal for the output
gain register of the CCD. We used frame rates between 18~Hz
and 140~Hz, with sub-array readout where necessary. The
CCD temperature was raised slightly on the second night (July 6) so as
to improve the charge transfer efficiency. Short exposure images taken
with high output register gain showed time-varying pattern noise with
an amplitude equivalent to $\sim$0.5 detected photons.

The f/11 beam at the focus was converted to f/60 using a single
achromat, giving an image scale of $27\times40$~milliarcsec per pixel
and a total imaging area of $11.5\times15.4$~arcsec.
In order to investigate the image quality achieved at large angular
separations from the reference star, the camera optics were designed
so that light from two regions of the sky separated by 25 arcsec
could be superimposed on the CCD. This allowed science targets to be
imaged at separations of up to 30~arcsec from the reference star.
All the observations were made at 810~nm with a top-hat filter of
120~nm bandwidth and with no autoguider in operation.

Good exposures were selected on the basis of
Strehl ratios calculated from each short exposure image of the
reference star, a technique demonstrated
previously in \cite{baldwin01}. The spurious bright spots found in
images taken
with our CCD65 detector (\cite{mackay01}) did not occur frequently
enough to have a significant effect on the reconstructed images,
although a small number of frames with bright spots close to the
reference star were excluded from our analyses.

The accuracy of measurement of the Strehl ratio and position of the 
brightest speckle was improved by convolving the individual
frames with the diffraction-limited telescope point-spread function
before these parameters were calculated. The peak
intensity in the convolved images represents the location where
correlation with the diffraction-limited point-spread function is maximised,
providing a very good estimate for the location and intensity of a
diffraction-limited speckle within the original image. After frames
have been selected based on the Strehl ratio for the brightest
speckle found, the corresponding unprocessed
exposures are sinc resampled, shifted and co-added to produce the
final image.

\section{Results}
\begin{figure}
\includegraphics[width=8.8cm]{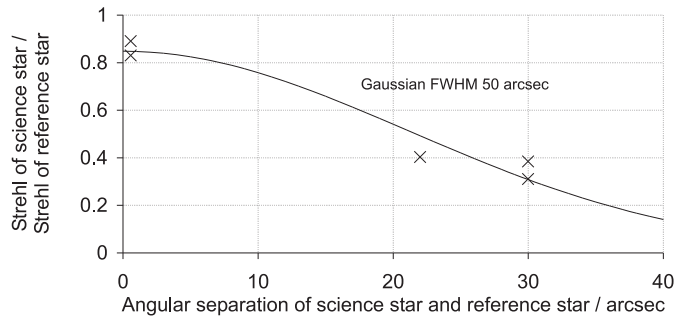}
\caption{
The Strehl ratio of each science object divided by the Strehl ratio for
the corresponding reference star is plotted against the angular
separation between science object and reference star. 
}
\label{isoplanatic}
\end{figure}

The size of the isoplanatic patch in Strehl-selected
exposures was deduced from observations of the binary stars HD 203991, 8 
Lac and 61 Cygnii, which have separations of 0.6, 22 and 30 arcsec 
respectively. These were all bright enough that the selection and
image shift correction for exposures was free of the effects of 
small number statistics in photon detection. One component of each 
binary was used as the reference to select the best 1\% of the exposures, 
for which both the image full width at half-maximum intensity (FWHM)
and Strehl ratio of the other component 
was also measured. Fig.~\ref{isoplanatic} shows the fractional reduction in
Strehl ratio of the second component as a function of angular separation 
from the reference star. 
A Gaussian fit to the points gives 50 arcsec for the FWHM of the 
isoplanatic patch for these selected frames, taken during a period in 
which the mean seeing was 0.51 arcsec. The image FWHM for objects 30
arcsec from the reference star using the best 1\% from a
run of 4000 exposures of 61 Cygnii
was 130 milliarcsec, only slightly poorer than the FWHM of $\sim$80
milliarcsec expected for a well sampled diffraction-limited
point-spread function. The measured FWHM is increased to 230~milliarcsec
when selecting the best 10\% of exposures, and 300~milliarcsec when selecting
all the exposures for a shift-and-add image. 

\begin{figure}
\includegraphics[width=8.8cm]{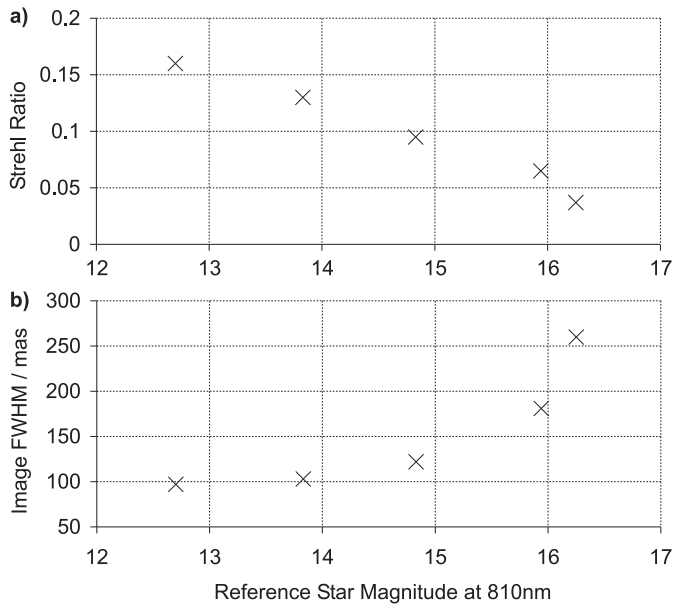}
\caption{
{\bf a, b.} Image resolution for a range of different reference stars in
M13.
}
{
\newcounter{alpha_count2}
\begin{list}{\textbf{\alph{alpha_count2})}}{\usecounter{alpha_count2}
	\setlength{\rightmargin}{\leftmargin}}
\item Typical Strehl ratios for stars within a few arcsec of the
reference star, for a range of reference star magnitudes.
\item Typical image FWHM in milliarcsec for these stars.
\end{list}
}
\label{limiting_mag}
\end{figure}

\begin{figure}
\includegraphics[width=8.8cm]{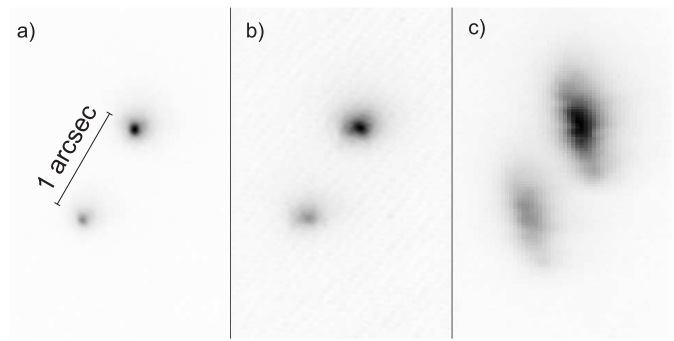}
\caption{
{\bf a-c.} Image resolution in M13 with different reference stars.  
}
{
\newcounter{alpha_count3}
\begin{list}{\textbf{\alph{alpha_count3})}}{\usecounter{alpha_count3}
	\setlength{\rightmargin}{\leftmargin}}
\item Image selection, shifting and adding using an $I=13.8$ reference
star 2.5 arcsec away. Image FWHM=100 milliarcsec, Strehl ratio=0.13
\item Image selection, shifting and adding using an $I=15.9$ reference
star 2.9 arcsec away. Image FWHM=180 milliarcsec, Strehl ratio=0.065
\item Conventional image of M13 produced by summing together 1000 of
the raw frames without shifting to compensate for image motion. Image
FWHM=$570\times390$ milliarcsec, Strehl ratio=0.019
\end{list}
}
\label{image_quality_comparison}
\end{figure}

The effect of reference star magnitude on the quality of the resulting 
image was studied using observations with the full field of view of the 
CCD in a densely populated region of M13. The 3.4~MHz pixel rate of the 
CCD controller limited the frame
rate for these observations to 18~Hz, allowing image motion to blur
the exposures.
This limited the Strehl ratios for reconstructed images to $\le$0.16,
with image FWHM of $\sim$100 milliarcsec. Data was taken from 6
runs of 1000 frames each.
The selection of the best
1\% of the 6000 exposures worked successfully
with reference stars as faint as $I=15.9$. The effects
of ``over-resolution'' on faint reference stars (\cite{nieto91}) were
avoided by using other objects in the field for image quality
measurements. Fig.~\ref{limiting_mag} shows the
variation in the Strehl ratio and FWHM of nearby stars when a range of different stars
are used as the reference for image selection, shifting and
adding. I band stellar magnitudes were taken from \cite{cohen97}.
The Strehl ratios with reference stars of $I=13.8$ and
$I=15.9$ are 0.13 and 0.065 respectively, a substantial improvement over
$\sim$0.019 for conventional astronomical images generated from
each of the individual runs. Fig.~\ref{image_quality_comparison} shows a
comparison of the image quality in these cases. The elongation in
Fig.~\ref{image_quality_comparison}c may be due to image drift over
the 55~s unguided exposure.

Given a suitable reference star, what combination of exposures gives the
best image? The simple method applied so far is to select the best $x$\%, $x$
being chosen to balance good Strehl ratio against adequate sensitivity. 
If the noise is Gaussian, a better solution would be to weight each
exposure proportionally to the square of the signal-to-noise ratio (snr)
for the quantity of interest. For example, detection of a faint
stellar image with Strehl ratio $S$ against a uniform noise background
would require weighting $\propto S^{2}$.
Since the value of $S$ for a faint target is unmeasurable in a single 
exposure, how well does the Strehl ratio for the reference star, $S_{\mathrm{ref}}$, 
represent that for the target?

\begin{figure}
\includegraphics[width=8.8cm]{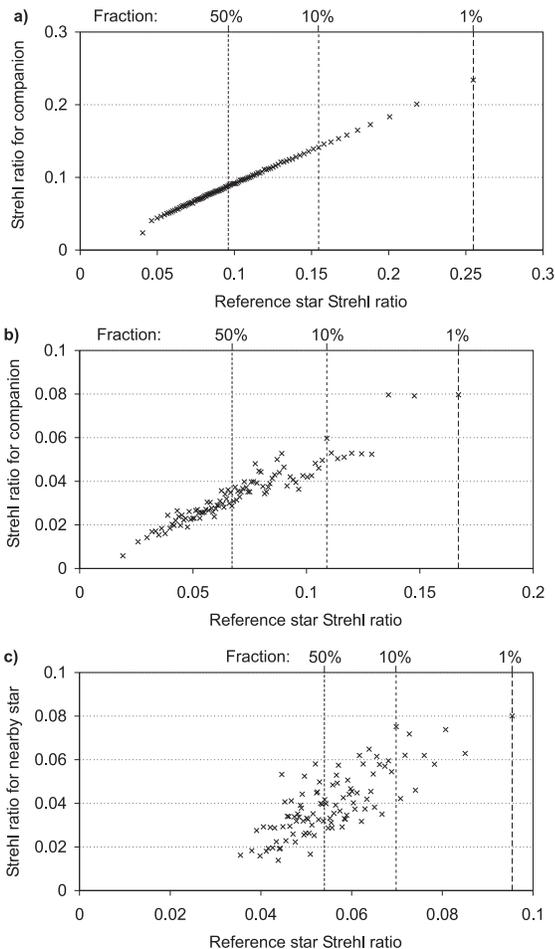}
\caption{
{\bf a-c.} Strehl ratios for reference star and field star in
shift-and-add images using selected short exposures.  
}
{
\newcounter{alpha_count4}
\begin{list}{\textbf{\alph{alpha_count4})}}{\usecounter{alpha_count4}
	\setlength{\rightmargin}{\leftmargin}}
\item Exposures of $\zeta$ Bo\"otis were sorted according to the
Strehl ratio measured on the A component (reference star) into 100
groups of 232 
exposures. The exposures in each group were shifted and co-added.
Fig.~\ref{science_vs_ref}a shows the resulting
Strehl ratio of the binary companion plotted against the reference
star Strehl ratio. The vertical lines
indicate the reference Strehls given by the best, the 10th best and the 50th
best group of exposures.
\item This process repeated for 4000 exposures taken on the 30 arcsec
binary star 61 Cygnii. 
\item 1000 exposures from M13 were analysed in the same way. The reference star had a
magnitude of $I=15.9$. Strehl ratios for a nearby $I=12.7$ star
are shown.
\end{list}
}
\label{science_vs_ref}
\end{figure}

To check this, data on the binary stars $\zeta$ Bo\"otis (from May
2000, \cite{baldwin01}), 61 Cygnii and a close pair of stars in M13
were reanalysed. The 
exposures in each run were ordered by $S_{\mathrm{ref}}$, placed in 100 equal bins   
starting with the best 1\%, then the next 1\% and so on. 
The exposures in each bin were shifted and co-added and the resulting 
Strehl ratio for the companion star was plotted against $S_{\mathrm{ref}}$
(Fig.~\ref{science_vs_ref}a-c).
All three plots show a linear relation between the Strehl 
ratios. The scatter of the points about the line increases from (a) to (c) 
mainly because of the decreasing number of exposures in each bin, but also 
in (c) due to photon statistics for the faint reference star. The latter 
also biases $S_{\mathrm{ref}}$, offsetting the mean line from the origin.
The interesting conclusion is that $S_{\mathrm{ref}}$ is a good measure of $S$ 
for the target objects, even for comparatively low Strehl ratios; there 
seems no lower limit at which the correlation collapses. In practice,
the limiting sensitivity using exposures weighted by $S_{\mathrm{ref}}^{2}$ was
determined by pattern noise produced by the camera
electronics. Exposure selection was found to give better rejection of
this pattern noise, and has been used for the remainder of our
analyses.

\begin{figure}
\includegraphics[width=8.8cm]{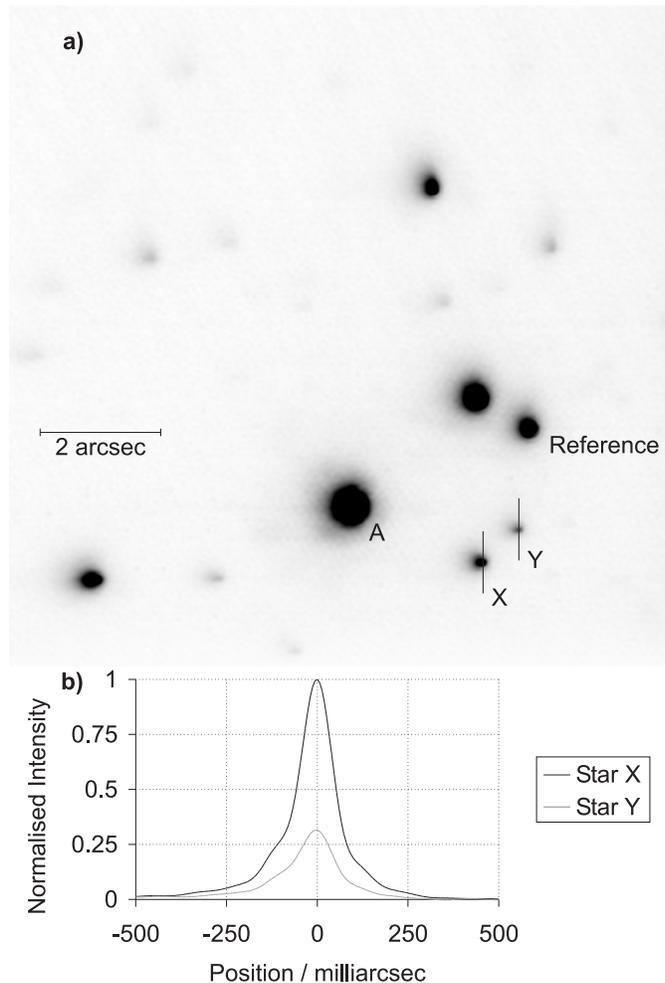}
\caption{
The best 10\% of exposures of M13 were selected, shifted and added
to produce image {\bf a)}. The reference star for frame selection and
recentring has been labelled, as has star {\emph A} from
\cite{cohen97}. North is at the top. Cross-sections through the stars
labelled X and Y are shown in part {\bf b)}.
}
\label{M13_lucky}
\end{figure}
Fig.~\ref{M13_lucky}a shows a high resolution image of a field near
the core of M13 generated
by selecting the 10\% of exposures with the highest Strehl ratios from
6000 taken at 18~Hz frame rate. The $I=12.7$ reference star
used for exposure selection and recentring has been labelled. Typical
stellar FWHM in this image are $\sim$120
millarcsec. Slight asymmetry in the faint halos around stars may relate to
reduced charge transfer efficiency on the CCD at low signal
level. Cross-sections through the stars with magnitudes 13.8 and 14.9
labelled X and Y are shown in Fig.~\ref{M13_lucky}b.

In order to asses the image quality, the 60 best exposures of
M13 were separated into two groups of 30. The
exposures in each group were shifted and added together giving two
independent images of the field in M13. These images were then
convolved with the diffraction-limited telescope point-spread function. Measurements
were made of the location and intensity of the brightest pixel for
each star, providing crude relative astrometry and photometry within the
images. The rms difference in astrometry between the two independent
datasets was found to be 6~milliarcsec, and the rms difference in
stellar magnitudes was 0.02. The limiting magnitude for a 5$\sigma$
detection of a star above the noise in these images was $I\simeq19$.

\section{Conclusions}
The observations described here show that: (i) the selection of short
exposure images can provide high resolution imaging up to 30 arcsec from a
reference star; (ii) Strehl ratios and image shift can be accurately
calculated with reference stars as faint as $I=15.9$; and (iii) assuming a mean
density of 400 stars per square degree at the South Galactic Pole with $I\le15.9$
(\cite{reid82}), 8\% of the South Galactic Pole region is within range
of a suitable reference star. If the mean stellar density over the
whole sky is 3.3 times higher, then $\sim$20\% of the sky
will be within range of suitable reference stars with our current
detector. The recent development of thinned low-noise CCDs and
improvements to our camera electronics
should improve the limiting reference star magnitude,
allowing high resolution imaging over $\ge$25\% of the sky. The
faintest point source detected at a
5$\sigma$ level from 10\% of exposures selected in an hour of
observing would be
$I\ge21$ with our old camera or $I\sim23$ with the improved
instrument.
\begin{acknowledgements}
The Nordic Optical Telescope is operated on the island of La Palma
jointly by Denmark, Finland, Iceland, Norway and Sweden, in the
Spanish Observatorio del Roque de los Muchachos of the Instituto de
Astrofisica de Canarias.
\end{acknowledgements}

\end{document}